\documentclass[nofootinbib,floats,aps,groupedaddress,preprint]{revtex4-1}
\usepackage{amssymb,amsmath,amsbsy}
\usepackage{mathrsfs,verbatim,enumerate}
\usepackage[dvips]{graphics}
\usepackage{epsfig}

\def\beq{\begin{equation}}
\def\eeq{\end{equation}}
\def\beqa{\begin{eqnarray}}
\def\eeqa{\end{eqnarray}}
\def\ifmath#1{\relax\ifmmode #1\else $#1$\fi}
\def\lsup#1{^{\lower 4pt\hbox{$\scriptstyle#1$}}}
\def\llsup#1{^{\lower 2pt\hbox{$\scriptstyle#1$}}}

\def\lsim{\mathrel{\raise.3ex\hbox{$<$\kern-.75em\lower1ex\hbox{$\sim$}}}}
\def\gsim{\mathrel{\raise.3ex\hbox{$>$\kern-.75em\low
er1ex\hbox{$\sim$}}}}

\def\fig#1{Fig.~\ref{#1}}

\newcommand{ \gravitino} {\mbox{$\tilde{G}$}}
\newcommand{ \nlsp }{\mbox{$\tilde{\chi}_{1}^{0}$}}
\newcommand{ \tauchione }{\mbox{$\tau_{\tilde{\chi}_{1}^{0}}$}}
\newcommand{ \chioneggrav } {\mbox{\nlsp $\rightarrow \gamma + \gravitino$}}
\newcommand{ \myhiggs } {\mbox{$h^0$}}
\newcommand{ \higgschichi } {\mbox{\myhiggs $\rightarrow$ \nlsp\nlsp}}
\newcommand{ \mhiggs } {\mbox{$m_{\myhiggs}$}}
\newcommand{ \mchi } {\mbox{$m_{\nlsp}$}}
\newcommand{ \gdelayed } {\mbox{$\gamma_{\rm delayed}$}}
\newcommand{ \tcorr } {\mbox{$t_{\rm corr}$}} 

\newcommand{ \basehiggs } {\mbox{\mhiggs = 135\mgev}} 
\newcommand{ \basechi } {\mbox{\mchi = 55.5\mgev}} 
\newcommand{ \basetau } {\mbox{\tauchione = 5~ns}} 
\newcommand{ \baserelation } {\mbox{\mhiggs = 2~$\cdot$ \mchi\ + 24~GeV}} 

\newcommand{\mgev}{\mbox{~${\rm GeV}/c^2$}}
\newcommand{\invfb}{\mbox{${\rm fb}^{-1}$}}
\newcommand{\PYTHIA}{{\sc pythia}}

\newcommand{\ET}{\mbox{$E_T$}}
\newcommand{\met}{\mbox{${E\!\!\!\!/_T}$}}
\newcommand{\Met} {\met}
\newcommand{\etg}{\mbox{$E_T^{\gamma}$}}

\begin{document}
\pagestyle{empty}
\begin{center}
{\Large\bf Prospects of Searches for Gauge Mediated Supersymmetry with \myhiggs $\rightarrow$ \nlsp \nlsp\ production in the Time-Delayed Photon + \Met\ Final State at the Tevatron}

\vspace{1cm}

{\sc John D. Mason$^{a,}$}\footnote{E-mail: jdmason@physics.harvard.edu}
{\sc\small and}
{\sc David Toback$^{a,}$}\footnote{E-mail: toback@tamu.edu}
\vspace{0.5cm}

{\it\small $^a$Harvard Center for the Laws of Nature, Harvard University,
Cambridge, Massachusetts 02138, USA}

{\it\small $^b$ Mitchell Institute for Fundamental Physics and Astronomy, Texas A\&M University, College Station TX 77843-4242}

\end{center}

\vspace{0.5cm}
\begin{center}
\today
\vspace{1cm}

\begin{abstract}

We propose a search for direct production and decay of the lightest supersymmetric Higgs boson to two neutralinos in gauge mediated models at the Fermilab Tevatron. We focus on the final state where each neutralino decays to photon and light gravitino with a lifetime of order $\mathcal{O}({\rm ns})$. In the detector this will show up as a photon with a time-delayed signature and missing \ET. We estimate that using the photon timing system at CDF, and the full 10~\invfb\ data sample, that the sensitivity can be within a factor of three in some regions of parameter space for direct production of the Higgs.

\end{abstract}

\end{center}
\maketitle

\newpage
\setcounter{page}{1}
\setcounter{footnote}{0}
\pagestyle{plain}

\section{Introduction and Overview}
The Higgs potential in the standard model (SM) provides a simple description of the dynamics of electroweak symmetry breaking. An explanation of why the electroweak scale is hierarchically smaller than the Planck scale is provided by embedding the Higgs potential in the minimal supersymmetric standard model (MSSM). The MSSM predicts the existence of a
variety of new supersymmetric (SUSY) particles. If SUSY breaking is communicated to the MSSM via gauge interactions~\cite{Dine:1981za,Dimopoulos:1981au,Dine:1981gu,Nappi:1982hm,
AlvarezGaume:1981wy,Dimopoulos:1982gm}, so-called gauge mediation supersymmetry breaking (GMSB), it is possible that the fundamental scale of SUSY breaking can be low, $\mathcal{O}(100~{\rm TeV})$, in which case the messenger and SUSY breaking scales are similar in magnitude ~\cite{Dine:1993yw, Dine:1994vc,Dine:1995ag}. In the most general framework~\cite{Meade:2008wd, Carpenter:2008wi,Buican:2008ws,Komargodski:2008ax,Rajaraman:2009ga, Intriligator:2010be,Dumitrescu:2010ha}, so-called general gauge mediation (GGM), a variety of superpartner spectra are possible. Since many explicit models fall into this broader class, it is important to consider them. In particular, GGM allows for the lightest and next-to-lightest sparticles to be the gravitino (\gravitino) and lightest neutralino (\nlsp) respectively and have masses less than 1 ${\rm keV}/c^2$ and 50\mgev , respectively. In the case that the \nlsp\ mass is near or below $M_{Z^0}$ we expect $BR(\chioneggrav)\approx$100\%. These scenarios lead to interesting $\gamma$ + \met\ final states if sparticles are produced at colliders~\cite{Dimopoulos:1996vz, Ambrosanio:1996zr,Ambrosanio:1996jn,Dimopoulos:1996va, Dimopoulos:1996yq, Lopez:1996gd,Baer:1996hx,Ambrosanio:1997rv,Kawagoe:2003jv,Shirai:2009kn,Meade:2009qv,Meade:2010ji}.

Current experimental results from searches for GMSB at LEP, the Tevatron and the LHC~\cite{LEPGMSB,TeVGMSB,CDFLongLived,LHCGMSB} are not sensitive to scenarios where the \nlsp\ and \gravitino\ are the only sparticles with masses that are kinematically accessible. The standard GMSB searches are focused on minimal gauge mediation (MGM) which is typically encapsulated using the SPS-8 relations~\cite{SPS8}, and often assume the lifetime of the \nlsp\ (\tauchione) is $\ll$1~ns. The reason most experiments are sensitive to MGM models is that they allow for the production of the heavier sparticles at a high rate, each of which decay down to \nlsp\ -pairs which in turn decay to $\gamma\gamma+\Met$ in association with other high energy SM particles. For low lifetimes, both photons can be observed in the detector as they are promptly produced. Other searches that assume a nanosecond or longer \tauchione, favored when the SUSY breaking scale is low~\cite{Dimopoulos:1996vz}, have also been done at both LEP~\cite{LEPGMSB} and the Tevatron~\cite{CDFLongLived}, but they also assume SPS-8 type relations, which keep the production cross sections high. If the mass relationships in SPS-8 are released, then it is possible that only the \nlsp\ and \gravitino\ have masses low enough to be kinematically allowed in collider experiments and the large direct sparticle production rate previously considered essentially vanish. In this case then the LEP, Tevatron and LHC limits no longer cover the low mass \nlsp\ scenarios~\cite{PromptHiggsGMSB}. We will refer to this case as the Light Neutralino and Gravitino (LNG) scenario. 

In this paper we discuss the potential for sensitivity to LNG models by focusing on the production of the lightest Supersymmetric Higgs (\myhiggs) at the Tevatron and its subsequent decay to \nlsp -pairs. If we consider the Electroweak fits and SUSY favored mass region 115\mgev\ $<$ \mhiggs\ $<$ 160\mgev~\cite{Inoue:1982ej,Carena:2002es} and assume a favorable mass relationship between the \myhiggs\ and the \nlsp, \mbox{(\mhiggs~$\geq 2~\cdot$ \mchi)},  then the production cross section for photon+\Met\ final states via $p{\bar p}\rightarrow \higgschichi \rightarrow (\gamma\gravitino)(\gamma\gravitino)$  can be in the picobarn range at the Tevatron and be a factor of~1,000 over production that proceeds via $Z^*/\gamma$ diagrams~\cite{PromptHiggsGMSB}. While single Higgs production is always a challenge because it will not produce many final state particles, the long-lifetime of the \nlsp\ can provide a smoking gun signature of exclusive photon+ \Met\, with a delayed arrival time of the photon at the calorimeter. Using the CDF photon timing system~\cite{EMTNIM} and the techniques in~\cite{CDFLongLived} we can identify these delayed photons, \gdelayed.

We propose a search for exclusive production of $p{\bar p}\rightarrow\higgschichi  \rightarrow$ \gdelayed+ \Met; the so-called exclusive \gdelayed+ \Met\ final state. We take advantage of the high production cross section of the \myhiggs, the nanosecond lifetime of the \nlsp, old phenomenology methods/results for $p{\bar p}\rightarrow Z^*/\gamma \rightarrow \nlsp\nlsp\ \rightarrow$ \gdelayed+ \Met~\cite{TobackWagner}, as well as improvements in the understanding of the EMTiming system at CDF~\cite{EMTNIM} and the SM backgrounds to the exclusive \gdelayed+ \Met\ final state searches~\cite{CDFLED}. As we will see, this search opens the exciting possibility of a simultaneous discovery of both the Higgs and low-scale SUSY using the full 10 \invfb\ data set at the Tevatron.

The outline of the paper is as follows, in Section~\ref{Theory} we briefly describe both the framework we consider for general forms of gauge mediation as well as how it couples to the Higgs sector. We then describe the assumptions and the experimental results used to constrain the parameter space we consider in the search. As we will see, sparticle production rates are well described by \mhiggs\ and its branching ratio to \nlsp-pairs. In Section~\ref{Analysis} we describe the analysis methods. Using existing tools and data, as well as simple analysis assumptions, we find that the sensitivity is determined solely by \mhiggs, \mchi\ and \tauchione. Each plays an important role in the kinematics of the events and the amount of delay of the photon. In Section~\ref{Results} we give our results, and in Section~\ref{Conclusion} we conclude that even with our simple assumptions, and no particular optimization, that we are within a factor of three of being sensitive to single Higgs production in a scenario that no one else is sensitive to.

\section{Gauge Mediated Supersymmetry and the Higgs Sector}\label{Theory}

%This about the Mechanism of symmetry breaking: Using  instead of MGM
We will consider the LNG scenario where only the \nlsp\ and the \gravitino\ have masses that are accessible at the Tevatron, as is allowed in GGM scenarios~\cite{Meade:2008wd} and not excluded by current searches for GMSB. In GGM models the bino mass ($M_1$), wino mass ($M_2$), and gluino mass ($M_3$) are free parameters. By way of contrast, in MGM models there is a rigid relation, $\frac{M_2}{M_1} \sim (\frac{g}{g'})^2$, where $g$ and $g'$ are the $SU(2)_L$ and $U(1)_Y$ gauge coupling strengths. This forces the chargino to be light if the \nlsp\ is also; current limits would imply \mchi $>$ 150\mgev. Since there is no reason these relationships must hold in Nature, a lighter \nlsp\ can easily be achieved in this context due to a general soft mass spectrum for superpartners which still preserves the flavor-blind mechanism of communicating SUSY breaking to the MSSM. In this case, it is possible that \mchi\ is of the order 50\mgev, the gravitino is less than a ${\rm keV}/c^2$, and all other sparticles are too heavy to be produced at the LEP, Tevatron or the LHC. Exclusive searches at LEP~\cite{LEPGMSB} can place very restrictive limits on a low mass \nlsp\, but are only applicable for situations with large direct \nlsp-pair production cross sections as in MGM models~\cite{PromptHiggsGMSB} which do not occur in this scenario.

%Current searches at LEP~\cite{LEPGMSB}, the Tevatron~\cite{TeVGMSB} and LHC~\cite{LHCGMSB} have focused on Minimal Gauge Mediated (MGM) %models (e.g. SPS-8~\cite{SPS8}) which typically include direct production of sparticles that cascade to the \nlsp\ which, in turn, %decays via \chioneggrav\. The branching ratio of \chioneggrav\ is $\approx 100\%$ to a very good approximation for \mchi $< %M_{Z^0}$~\cite{GammaGrav100}. 

%This is about the Higgs and mass region for the Higgs that we look in. 
In addition to the sparticle spectrum of the MSSM, the two-higgs doublets provide five separate physical Higgs particles. We note that most of SUSY parameter space is such that the \myhiggs\ is SM-like in its couplings to SM particles. For this reason, it is reasonable to work in the decoupling limit \cite{Gunion:2002zf}. Furthermore, if 2$\cdot$\mchi $<$ $\mhiggs$ the branching fraction of the Higgs to \nlsp\ pairs, $BR(\higgschichi)$, can become significant. In this case, the LEP and Tevatron bounds on the SM Higgs mass are applicable to \mhiggs, but must be modified in order to take into account the inclusion of the \higgschichi\ decay mode. The SM Higgs mass bound from LEP,  $m_{\rm Higgs} >$ 114.4\mgev\ at $95\%~{\rm C.L.}$~\cite{Barate:2003sz}, is only slightly modified by the inclusion of our new decay process since, as we will show, $BR(\higgschichi) < 0.7$. The Tevatron 95\% C.L. exclusion region for the SM Higgs, 153\mgev\ $< m_{\rm Higgs} <$173\mgev~\cite{TeVHiggsLimits}, will also be slightly reduced. For the scope of this paper we consider the Higgs mass as a free parameter~\cite{HiggsNote} and consider the range 120\mgev\ $<$ \mhiggs $<$ 160\mgev, favored by electroweak fits. For this mass region, the Higgs's production cross-section is dominated by the $gg$ fusion diagram and is in the picobarn range and effectively determined by \mhiggs\ alone~\cite{HIGLU}. 

%%%%%%%%%%%%%%%%%%%%%%%%%%%%%%%%%%%%%%%%%

\begin{figure}[htb]
\begin{tabular}{cc}
\includegraphics[scale=0.9]{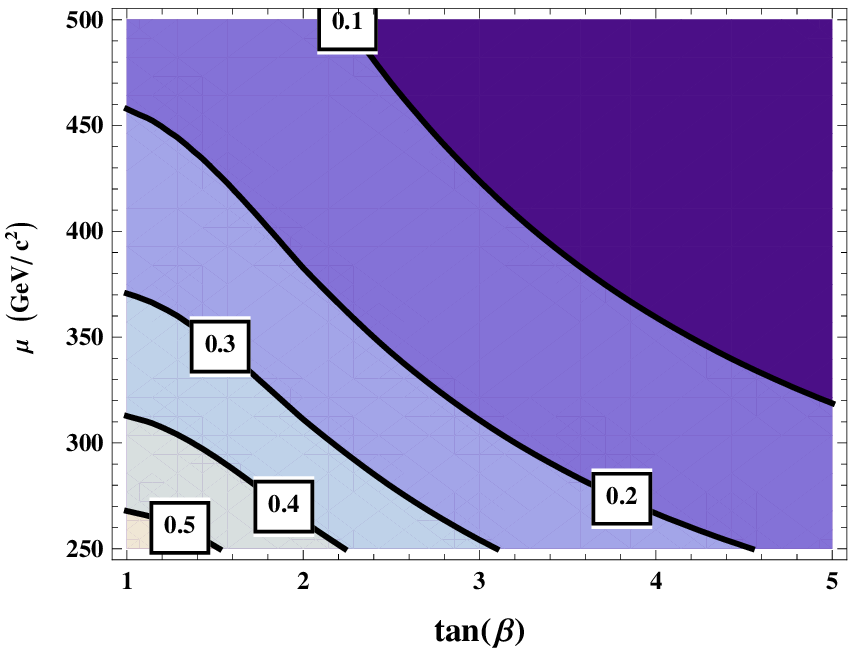} & \includegraphics[scale=0.9]{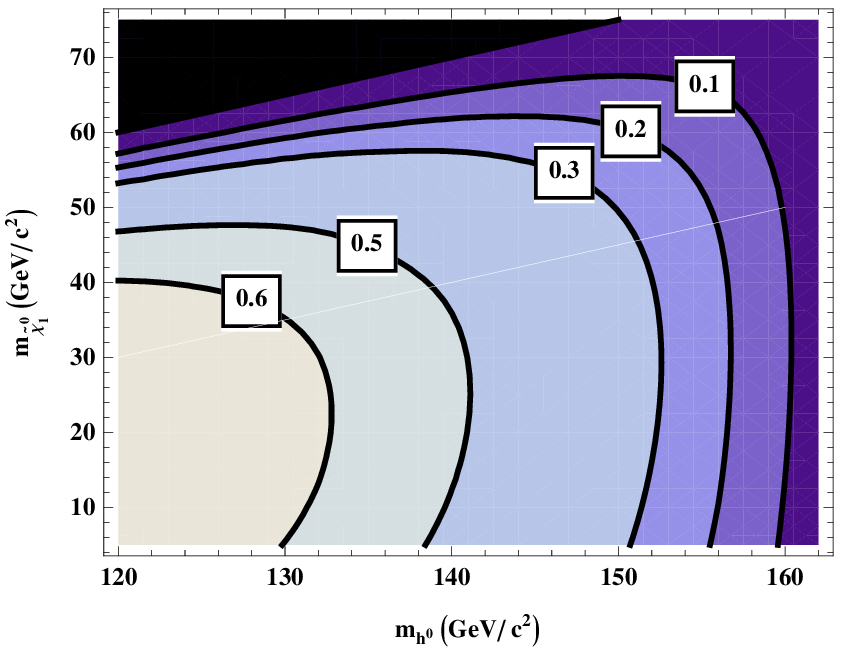} \\
\end{tabular}
\caption{The contours of constant branching fraction of \higgschichi. On the left shows the results in the  $\mu$-$\tan{\beta}$ plane with \basehiggs\ and \basechi.  On the right is the results in the \mchi-\mhiggs\ plane with $\tan{\beta} = 1.5$ and $\mu = 300 \mgev$. We note that the black region is kinematically forbidden. }
\label{brstuff}
\end{figure}

% This is the paragraph about the branching fraction to chi10 pairs. 
The branching fraction $BR(\higgschichi)$ can often be as large as 50\%. The width of \higgschichi\ is determined by \mhiggs\ as well the full widths to other modes such as $\bar{b}b$ and $W^+W^-$, if kinematically accessible, and the values $M_1, M_2$, $\tan{\beta}$, and $\mu$. We note that $\tan{\beta}$ and $\mu$, like $M_1$ and $M_2$, are independent parameters in GGM \cite{Komargodski:2008ax}. As long as the other superparticles remain kinematically inaccessible at the Tevatron, only these four parameters affect the Higgs branching ratio. The $BR(\higgschichi)$ is largest for small values of $\tan{\beta}$ and $\mu$ and fall as either $\tan{\beta}$ or $\mu$ grow. Results for $BR(\higgschichi)$ are shown in Figure~\ref{brstuff} for \mhiggs\ $= 135~{\rm GeV}$ and \mchi\ $= 55~{\rm GeV}$ but as a function of $\tan{\beta}$ and $\mu$, and can be bigger than 50\%. Since these parameters do not significantly affect other properties of the \myhiggs, \nlsp\ or \gravitino\ the values of $\tan{\beta}$ and $\mu$ can be thought of as being implicit in a choice of $BR(\higgschichi)$. While the BR is fairly insensitive to $M_2$, it is sensitive to \mchi\ and \mhiggs\, also shown in Figure \ref{brstuff}.  This result, in conjunction with the large \myhiggs\ production cross section shows why the production and decay of \higgschichi\ at the Tevatron can easily be a thousand times larger than $p{\bar p}\rightarrow Z^*/\gamma \rightarrow \nlsp\nlsp$ for appropriate choices of \mhiggs~\cite{HIGLU}. 

The final state phenomenology of \higgschichi\ is very different than that produced in SPS-8 scenarios~\cite{SPS8}. In the LNG scenario, spartice production is dominated by \myhiggs\ events which yields \nlsp-pairs; in SPS-8 \nlsp-pairs are produced at the end of decay chains, and thus are associated with large amounts of high energy final state particles from the cascades which makes them easier to separate from SM backgrounds. While $W^{\pm}\myhiggs$ and $Z^0\myhiggs$ processes can occur, their rate will be much smaller. New discovery methods will be needed at the Tevatron.

%%%%%%%%%%%%%%%%%%%%%%%%%%%%%%%%%%%%%%%%%%%%

% Fourth paragraph is about the chi10 lifetime
A crucial issue for any new search in LNG scenarios is that \tauchione\ of order $\mathcal{O}({1\rm~ns})$ is favored for models with a low fundamental scale of SUSY breaking. The \nlsp\ lifetime is given~\cite{Dimopoulos:1996vz} by:
\beq
c\tauchione\ = 48\pi\,\frac{m_{3/2}^2\,M_{\rm Pl}^2}{\mchi^5}
\,\frac{1}{|P_{1\gamma}|^2},
\eeq
where $|P_{1\gamma}| = |N_{11}\,c_W+N_{12}\,s_W|$ and N is the unitary rotation that diagonalized the neutralino mass matrix ($\mathcal{M}^0_D =N^{-1}\mathcal{M}^0N$),
and $m_{3/2} = \frac{|F|}{\sqrt{3}M_{\rm Pl}}$. The value of F (or equivalently $m_{3/2}$) is related to the value of the superpartner masses through the dynamics of SUSY breaking. In known calculable models, the fundamental scale of SUSY breaking is typically bounded by: $16\pi^2\frac{m_{sq}}{g_3^2} <|F|^{1/2}$. For squark masses of $m_{sq} \sim 2~{\rm TeV}$ ($m_{sq} \sim 10~{\rm TeV}$) this bounds $ 200 ~{\rm TeV} <|F|^{1/2}$ ($ 1000 ~{\rm TeV} <|F|^{1/2}$), which corresponds to a lifetime range of 0.4~ns $<$ \tauchione\ $<$ 180~ns.

%%%%%%%%%%%%%%%%%%%%%%%%%%%%%%%%%%%%%%%%%%%%

% Fifth paragraph is about the phenomenology at the Tevatron 

Previous studies of GMSB phenomenology at the Tevatron indicate that even if only \nlsp-pairs can be produced at the Tevatron, different final states must be considered for the lifetime regimes \tauchione\ $\ll$ 1~ns, 1~ns~$<$~\tauchione\ $<$ 50~ns and \tauchione\ $>$ 50~ns~\cite{TobackWagner}. For \tauchione\ $\ll$ 1~ns the photons will be produced promptly. The prospects of searches for $p{\bar p}\rightarrow$\higgschichi $\rightarrow (\gamma\gravitino)(\gamma\gravitino) \rightarrow \gamma\gamma$+\Met\ are described in \cite{PromptHiggsGMSB} with corresponding versions from $Z^*/\gamma$ described in \cite{TobackWagner}. In the case \tauchione\ $>$ 50~ns, both \nlsp\ -pairs will leave the detector and SUSY is largely undetectable using direct methods at the Tevatron. In the case 1~ns~$<$ \tauchione\ $<$ 50~ns, the final cascade of \chioneggrav\ happens at a spatial location that is significantly displaced from the primary collision event that produced the \myhiggs. While this can produce the $\gamma\gamma$+\met, the $\gamma+\Met$ and the \met\ final states, in each case the arrival time of the photon can be delayed relative to expectations than if it were promptly produced. This is known as a delayed photon or $\gdelayed$.  As shown in~\cite{TobackWagner}, having a long-enough lifetime to produce a delayed photon also typically produces the case where a significant fraction of the events have one \nlsp\ escaping the detector entirely, making the \gdelayed+\Met\ final state more sensitive than \gdelayed\gdelayed+\Met\ \cite{TobackWagner}. Finally, since only \nlsp-pairs are produced we must search in the exclusive \gdelayed + \Met\ final state. While there has been a search for long-lived \nlsp$\rightarrow \gamma\gravitino$ at the Tevatron using the delayed photon final state~\cite{CDFLongLived}, there is no Tevatron analysis of exclusive \gdelayed+\Met\ final state. LEP has performed a search in exclusive \gdelayed+\Met\ \cite{LEPGMSB}, but in the LNG scenario at LEP the production rates of sparticles would be negligible.

%Summary
To summarize, we have outlined an important and uncovered scenario where only the \nlsp\ and \gravitino\ are kinematically accessible at the Tevatron in GGM models.  In this model, production of \myhiggs\  $\rightarrow$ \nlsp \nlsp\  can be large and is well described by \mhiggs\ and $BR(\higgschichi)$ alone. For the favored nanosecond lifetime region, we expect the best sensitivity to be in $p{\bar p}\rightarrow \myhiggs \rightarrow \nlsp \nlsp \rightarrow (\gamma\gravitino)(\gamma\gravitino) \rightarrow$ exclusive \gdelayed+ $\Met$ if the Higgs is more that twice as heavy as \nlsp. We next turn to the sensitivity of this search which, as we will see, is dependent on \mhiggs, \mchi, and \tauchione. For the reasons above we considered a number of different mass and lifetime combinations in the phenomenologically favored regions: 120\mgev\ $<$ \mhiggs $<$ 160\mgev, 30\mgev\ $<$ \mchi $<$ 80\mgev\ and $1~{\rm~ns} < \tauchione < 20~{\rm~ns}$. For simplicity we choose a baseline scenario with \basehiggs, \basechi\ and \basetau\ since it is near the central values of our parameters.

\section{Analysis}\label{Analysis}

Our proposal is to use the photon timing system at CDF to search for an excess of exclusive \gdelayed + \Met\ events above background expectations with the full Tevatron dataset of 10~\invfb. A similar idea was proposed in 2004~\cite{TobackWagner}, but was based on the kinematics of \nlsp-pair production through $Z^*/\gamma$ and only crude analysis methods were employed. Since then the CDF EMTiming system has been installed and commissioned~\cite{EMTNIM}, delayed photon searches have been shown to be viable at the Tevatron~\cite{CDFLongLived} and CDF has completed a sophisticated Run~II version of the search for $\gamma$ + \Met\ events with a jet veto to enforce the exclusive final state, but without the timing requirement~\cite{CDFLED}. While we are sure that any actual exclusive \gdelayed+\Met\ search will be more sophisticated than what we are proposing, we use the current published results as well as Monte Carlo (MC) simulation methods to reliably estimate a sensitivity. We will use simple requirements to define our signal regions to estimate the backgrounds and acceptance to the search. Our primary emphasis is on robustness, so we will not introduce additional requirements where we cannot confidently model the backgrounds. 

The estimate of our sensitivity requires a number of elements. This includes the expected production cross sections, the branching ratios, the backgrounds for the proposed cuts (with associated uncertainty), the acceptances for the signal (with associated uncertainty), and the luminosity. For simplicity, we define the sensitivity as the expected 95\% confidence level (C.L.) cross section times branching ratio upper limit in the no-signal assumption scenario~\cite{Boos}. This allows for a comparison to various production cross section predictions that can be model or parameter choice dependent. We also choose to make our predictions based on the results of a straight-forward counting experiment where we compare the number of events in a signal region to background expectations as these are readily converted into an expected cross section limit. Thus, to estimate the sensitivity we simply require an estimate of the number of background events that pass all the final event selection requirements and the acceptance, which we define to be the fraction of the $\myhiggs \rightarrow \nlsp \nlsp\ \rightarrow  (\gamma\gravitino)(\gamma\gravitino)$ events passing those same requirements. In addition, we take into account some reasonable expectations for uncertainties as well as assume the full Tevatron run dataset with a 6\% luminosity uncertainty. For the event selection requirements we will use a combination of selection requirements from the published CDF papers.

We walk through these elements systematically. Since we assume the \myhiggs\ is SM-like in its couplings to SM particles, its production cross section is the same as for the SM Higgs and is determined solely by \mhiggs. The largest production mechanism of a Higgs is through $gg \rightarrow h^0$ and is the only one we will consider. This production cross section receives large enhancements from radiative corrections at NLO and are calculated using the HIGLU program~\cite{HIGLU}. NNLO corrections, as calculated in~\cite{NNLOXsec}, are incorporated using $k$-factors.

The background and acceptance estimates are based on a combination of published results and MC simulation. For simplicity, for the backgrounds we follow the available data from the CDF search for new physics in the exclusive $\gamma$+\Met\ final state in Ref.~\cite{CDFLED} and use these cuts as our baseline selection requirements. We then use a simple set of additional requirements.  The baseline requirements from the paper include a single isolated photon with $|\eta|<1.1$, $E_T > 40~{\rm GeV}$,  and the requirement of \Met~$>~$50 $~{\rm GeV}$. In addition, to reduce the large SM backgrounds, the event is rejected if there are any extra high energy objects in the event, such as an extra lepton or jet using a jet veto.  Since the data is well described as a function of $E_T$ in~\cite{CDFLED}, we can consider raising the $E_T$ requirement. Table~\ref{optcuts} lists the final set of requirements as well as the event reduction. We scale the results from 2~\invfb\ to 10~\invfb. 

Since the kinematics of the backgrounds are assumed to be independent of the timing of the photon~\cite{EMTNIM}, we consider them to be uncorrelated and follow the recommendations of~\cite{TobackWagner,CDFLongLived}. The photon timing variable at CDF compares the time of arrival of a photon candidate at the calorimeter relative to expectations. We define  ``\tcorr"  as,
\beq
\tcorr = (t_f-t_i) - \frac{|\vec{x}_f-\vec{x}_i|}{c},
\eeq
where $(t_i,\vec{x}_i)$ is the space-time location of the primary collision vertex and $(t_f,\vec{x}_f)$ is the space-time location of the photon when it deposits energy into the EM calorimeter. For a promptly produced photon with perfect measurements we would have \tcorr=0. Due to measurement uncertainties, for photons with a correctly identified vertex, the distribution is well described as  a Gaussian with a mean of zero, and an RMS of 0.65~ns~\cite{EMTNIM}. However, for this sample, where there is a jet veto, there are likely to be only a small amount of charged particles available to produce the vertex. In addition, the high luminosity running at the Tevatron is likely to produce multiple min-bias collisions which can be incorrectly selected as the vertex. This produces random values of $t_i$ and $\vec{x}_i$, where each is distributed according to the beam parameters which can each be described as a Gaussian with an RMS of $1.28 ~{\rm ns}$ and $28~{\rm cm}$ respectively. This scenario has been studied in~\cite{EMTNIM} which describes the ``wrong vertex" background as being well modeled as a Gaussian with a mean of zero, and an RMS of 2.05~ns. Because there is a high probability of picking the wrong vertex, we conservatively assume that 25\% of all background events will have an incorrectly assigned vertex. Putting this together, we take the background timing distribution to be uncorrelated with the kinematics of the event, double Gaussian, with both Gaussians's centered at zero, but 25\% having an RMS of 2.05~ns, and the rest with an RMS of 0.65~ns. The standard timing requirement is \tcorr$>$2~ns~\cite{CDFLongLived}, although this could, in principle, be optimized. This requirement rejects about 95\% of the backgrounds. We take an uncertainty on the final background estimate to be 30\%.

To estimate the acceptance for the signal we model $p{\bar p}\rightarrow \higgschichi \rightarrow (\gamma\gravitino)(\gamma\gravitino)$ using the \PYTHIA\ 6.4 MC event generator~\cite{PYTHIA} and the PGS4~\cite{PGS} detector simulation. We have modified PGS4 to recalculate the calorimeter cell in which a photon deposits energy for the case that the photon arises from the decay of a \nlsp. This properly takes into account the fact that the \nlsp\ decays at a position that can be different than location of the primary vertex. Similarly, we modified PGS4 to calculate the \tcorr\ for signal events. We measure the acceptance by counting the fraction of events that pass each of the final event-level reduction requirements in Table 1. To correct for the fact that we are comparing to the NNLO production, but using a LO MC simulation, we reduce the acceptance accordingly. Taking the ratio of the 0-jet production cross section, $\sigma^{(NLO)}_{0-jet}(gg \rightarrow h^0)$, to the $\geq$ 1-jet cross section we take an additional $75\%$ jet veto efficiency which we use in the final acceptance~\cite{NNLOXsec}. The final acceptance, as a function of the cuts, is displayed explicitly in Table~\ref{optcuts} for our baseline scenario. Following the recommendations of Ref.~\cite{TobackWagner,CDFLongLived} we assume a 20\% uncertainty on the acceptance.

%.   of the full NNLO production cross-section

\begin{table}[t!]
\centering
\begin{tabular}{|l||c|c|} \hline
Cut 				&  Signal Acceptance  & Background Events \\ \hline
$|\eta^{\gamma}| < 1.1$ 	&  $42\%$ 	& -\\ \hline 
\Met\ $>$ 50~GeV 		&  $1.7\%$ 	& - \\ \hline 
\etg\ $> 40$~GeV		&  $1.67\%$ 	& - \\ \hline 
Jet veto			&  $1.21\%$ 	& 2100\\ \hline 
\tcorr\ $>$ 2~ns 		&  $0.28\%$ 	& 89 \\ \hline 
\etg $>$ 50~GeV			&  $0.12\%$ 	& 52  \\ \hline  
\end{tabular}
\caption{\label{optcuts} Table of requirements to select exclusive \gdelayed+\Met\ events. In this table we assume \basehiggs, \basechi\ and \basetau. The acceptance is the fraction of events passing all the requirements, and takes into account the 75\% jet-veto efficiency starting in that row. We take a 20\% uncertainty on the acceptance.  The backgrounds are scaled to expectations for 10~\invfb\ and we assume a 30\% uncertainty.}
\end{table}

Given the  background, acceptance, luminosity and uncertainties we use a modification of the Corlim program~\cite{WalkerCorlim} in order to compute the expected 95\% C.L. cross section upper limit. In addition to the baseline selection requirements and the \tcorr $>$ 2~ns requirement, we found that raising the \etg\ to be \etg $>$ 50~{\rm GeV} was helpful. We considered raising the \tcorr\ and the \etg\ requirements further, but found either similar or lower sensitivity. Seeing no gain, we find a final background estimate of $52 \pm 16$ events. 

\section{Results}\label{Results}

\begin{figure}[htb]
\includegraphics{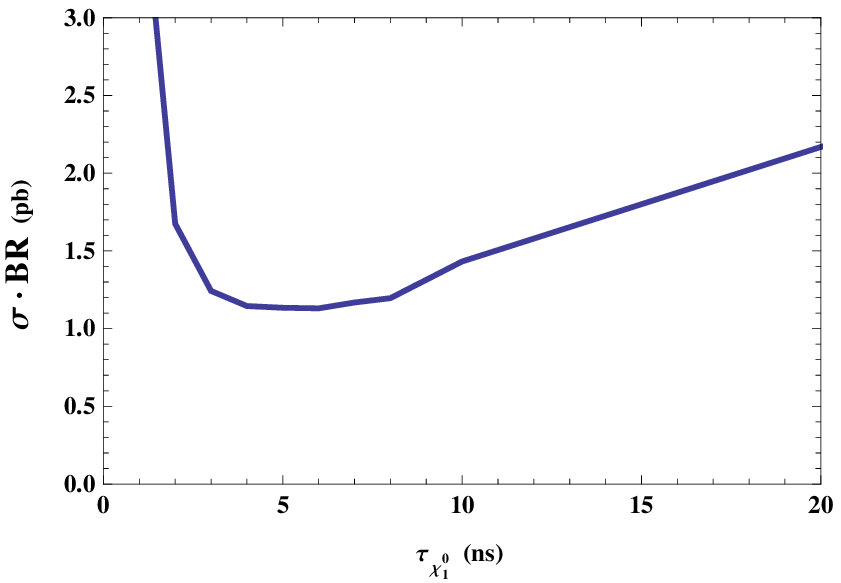} 
\caption{The expected 95\% C.L. cross section limit as a function of \tauchione, but where we have chosen \basehiggs\ and \basechi. }
\label{tau} 	
\end{figure}

We next consider the expected sensitivity as a function of  \mhiggs, \mchi\ and \tauchione. We begin by looking at the sensitivity as a function of \tauchione\ for fixed values of the \mhiggs\ and \mchi\ at their baseline values of \basehiggs\ and \basechi. The results are shown in Figure~\ref{tau}. We find that the optimal sensitivity occurs for a lifetime of 5~ns. This is readily understood in terms of kinematic arguments, and is consistent with the results of~\cite{TobackWagner,CDFLongLived}. For low lifetimes, \tauchione $\ll$1~ns, the \nlsp\ does not travel long enough within the detector to produce a \gdelayed\ with \tcorr\ $>$~2~ns. Said differently, the acceptance for the \tcorr$>$2~ns requirement goes to zero and the expected limit gets far worse. On the other side, as \tauchione\ gets large, for example \tauchione$>$10~ns, a larger and larger fraction of the \nlsp\ will leave the detector before decaying so the acceptance goes down as well.

\begin{figure}[htb]
\includegraphics{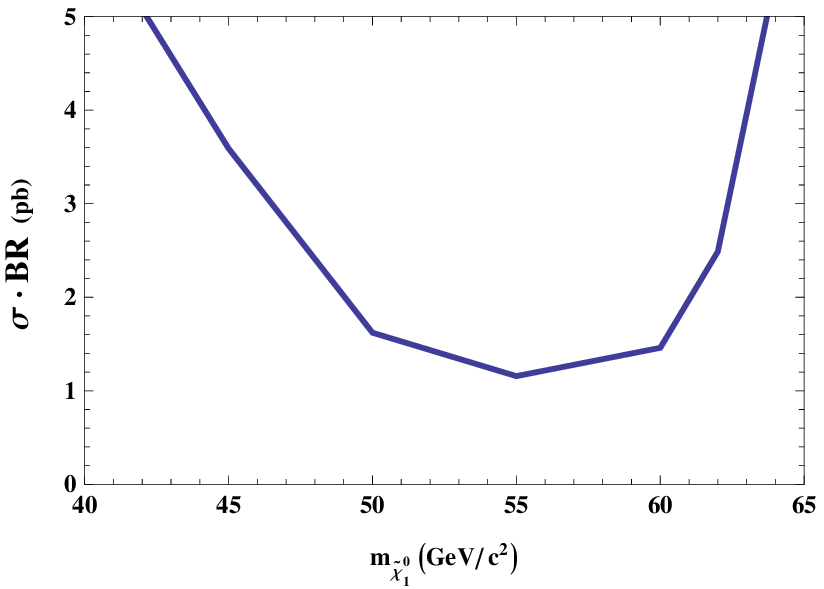} 
\caption{The expected 95\% C.L. cross section limit as a function of \mchi, but where we have fixed \basehiggs\ and \basetau. }
\label{limitvsmchi} 	
\end{figure}

\begin{figure}[htb]
\includegraphics{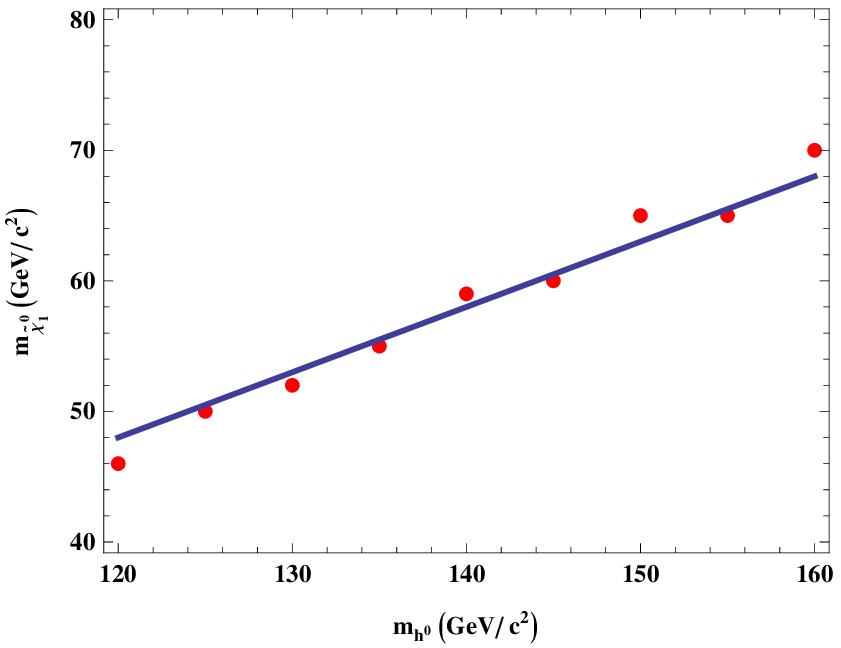} 
\caption{The value of \mchi\ that optimizes the sensitivity for a given \mhiggs. Here we have chosen \basetau. The line is given by:  $\mhiggs = 2 \cdot \mchi + (24 ~{\rm GeV})$.}
\label{optmhmchi} 	
\end{figure}

It is also useful to consider how the sensitivity varies as a function of \mhiggs\ and \mchi\ for a fixed \tauchione = 5~ns. The acceptance is sensitive to both masses individually, as well as in combination. In particular, the larger the value of \mhiggs\ the more energy there is available for the photon and \Met-producing objects to go above the selection requirement thresholds in Table~\ref{optcuts}. Similarly, the mass difference affects the kinematics as well. Equally important, as shown in Ref.~\cite{TobackWagner}, is the boost of the \nlsp\ which has an important effect on \tcorr\ since it also affects both the path length difference between the arrival position in the calorimeter and the original direction of the \nlsp. To study this latter variation we consider a \mhiggs\ fixed at its baseline value (and \basetau), and map out the sensitivity as a function of \mchi. This is shown in \fig{limitvsmchi}. The minimum of the distribution optimizes the expected limit. Repeating the results for all Higgs masses and picking the \mchi\ that minimizes the cross section, see Figure~\ref{optmhmchi}, we see a clear relationship which is well approximated by \baserelation. This optimal relationship arises for a number of reasons. On one side is the \etg\ $> 50~{\rm GeV}$ cut. Intuitively, the \myhiggs\ is mostly produced at rest in the lab frame and the \nlsp\ must carry some kinematic ``kick" so that the photon it emits has enough energy to pass these hard cuts. The lighter the \mhiggs\ the more of a kick the \nlsp\ needs. This favors small values of \mchi. However, if the \nlsp\ becomes very boosted, then the emitted photon travels in the same direction as the \nlsp\ in the lab frame and this reduces the value of \tcorr. This effect favors larger values of \mchi. The minimal value of the expected limit value reflects this balance.

\begin{figure}[htb]
\includegraphics{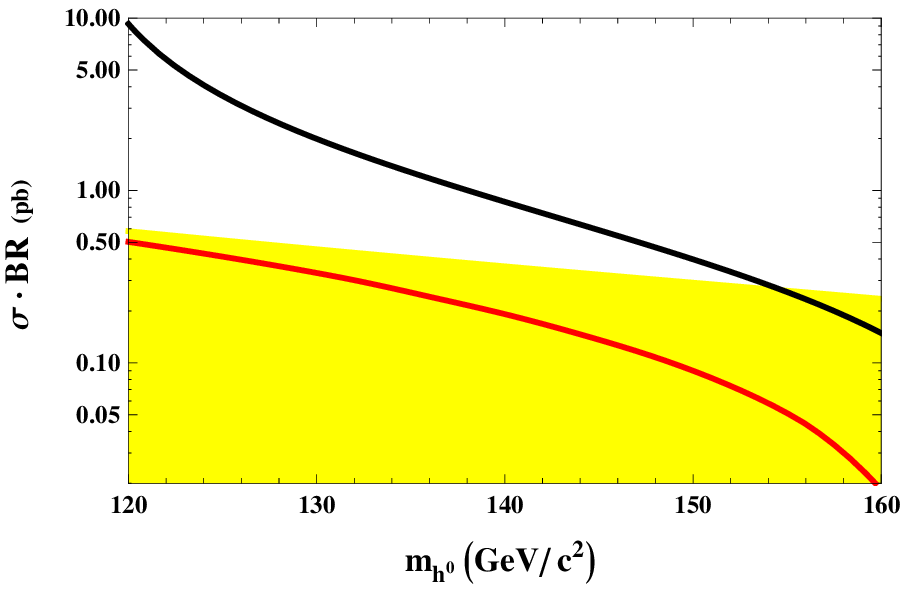} 
\caption{A comparison of the expected 95\% C.L. cross section limits where we have chosen \basetau\ and used the mass relation \baserelation\ for the calculation. The yellow band corresponds to BR = [0,0.5] with the top being the relation $\sigma(\myhiggs)\cdot 0.5$. The red line corresponds to $\sigma(\myhiggs)\cdot BR(\higgschichi)$ where the branching ratio is taken from the MSSM prediction for $\tan{\beta} = 1.5 $ and $\mu = 300\mgev$. }
\label{sum} 	
\end{figure}

%The solid blue line in the plot gives the Neutralino mass for which maximal possible photon $E_T$ is $50 \mgev$. Note that this is more or less aligned with the optimal relationship between \mhiggs\ and \mchi. The Neutralinos need to %be just light enough so that the photons that they decay to pass the hard $E_T$ cut.

% We can further illustrate the optimal relation as a contour plot of constant cross section \fig{contour}. { \bf probably will overlay previous plot on this contour}.  

To compare our expected cross section limits to the production cross section and branching ratio predictions we consider two branching ratio scenarios. For the expected limit on $\sigma \cdot BR (\higgschichi)$, we show the result as a function of \mhiggs\ where we have fixed \tauchione = 5~ns and taken \mchi\ according to \baserelation. This is shown in \fig{sum} as the black curve. Note that the expected sensitivity gets better and better for higher \mhiggs\ as more and more events pass can pass the kinematic thresholds. We first compare this to the expected $\sigma \cdot BR$ for  $\tan{\beta} = 1.5$ and $\mu = 300 \mgev$ (red line) where both $\sigma$ and $BR(\higgschichi)$ depend on \mhiggs. The second comparison is to the prediction for $\sigma \cdot BR$ where BR $\leq$ 0.5. This is shown as a yellow band. While our sensitivity is clearly dependent on \mhiggs, \mchi, \tauchione\ as well as $\mu$ and $\tan{\beta}$, we see our sensitivity is often within a factor of 5 of the expected production cross section times branching ratio. At some locations it is as close as a factor of 3.

\section{Conclusion}\label{Conclusion}

We have investigated the sensitivity of a proposed search in the exclusive \gdelayed + \Met\ final state at CDF for direct production and decay of \higgschichi\ in gauge mediated models at the Fermilab Tevatron. While we have picked a fairly restricted regime, in many ways we are picking the favored regions of parameter space, and regions which are not yet covered by existing experiments. We find that within these assumptions we have optimal sensitivity when \tauchione\ of the order of 5~ns and when the mass of the \nlsp\ is slightly less than half the mass of the \myhiggs. While it is possible to consider lower lifetime searches, \tauchione $\ll$ 1~ns~\cite{TobackWagner,PromptHiggsGMSB}, we note that similar searches have not found any evidence of new physics \cite{TeVGMSB}. We estimate that using the photon timing at CDF, and a data sample of 10~\invfb\, that the sensitivity can be within a factor of three of some regions of parameter space for direct production of the Higgs. 
\\

{\large {\bf Acknowledgements}} \\
 \\
The authors would like to thank the Harvard Center for the Fundamental Laws of Nature and the Mitchell Institute for Fundamental Physics and Astronomy at Texas A\&M University. We would like to thank Joel Walker for the use of his customized limit setting tools. DT would like to thank Jonathan Asaadi, Adam Aurisano,  Bhaskar Dutta, Daniel Goldin, Eunsin Lee, Jason Nett and Peter Wagner for helpful discussions. JDM would like to thank David Poland for useful discussions and would like to acknowledge the generous support of NSF Grant PHY-0855591, and the Aspen Center for Physics for their hospitality during the preparation of this manuscript

%\begin{figure}
%\includegraphics{mt_contour.eps} 
%\caption{ contours of cross section limit as a function of \mhiggs\ and \mchi. }
%\label{cont} 	
%\end{figure}

%\begin{figure}[htb]
%\includegraphics{mt_limitmhvsmchi.eps} 
%\caption{ contours of 95\% C.L. limit as a function of \mchi\ and \mhiggs}
%\label{contour} 	
%\end{figure}

\end{document}